\documentclass[onecolumn,showpacs,preprintnumbers,amsmath,amssymb,prb,preprint]{revtex4}

\usepackage{graphicx}
\usepackage{dcolumn}
\usepackage{bm}
\usepackage{color}

\begin{document}
\title{Coulomb interaction-induced Aharonov-Bohm oscillations}

\author{Toshihiro Kubo$^{1}$}
 \email{kubo.toshihiro.ft@u.tsukuba.ac.jp}
\author{Yasuhiro Tokura$^{1,2}$}%
 \affiliation{%
$^1$Graduate School of Pure and Applied Sciences, University of Tsukuba, Tsukuba, Ibaraki 305-8571, Japan\\
$^2$NTT Basic Research Laboratories, NTT Corporation, Atsugi-shi, Kanagawa 243-0198, Japan}%

\date{\today}

\begin{abstract}
We study the Coulomb interaction-induced Aharonov-Bohm (AB) oscillations in the linear response transport through a remote quantum dot which has no tunnel coupling but has Coulomb coupling with the quantum dot embedded in an AB interferometer. We show that the Coulomb interaction-induced AB effect is characterized by a charge susceptibility of a remote quantum dot in a weak interaction regime. In a strong but finite interaction regime, around the particle-hole symmetric point, there exists the region where the visibility of the induced AB oscillations becomes one although the visibility of the original AB oscillations in the interferometer is low.
\end{abstract}


\maketitle
\section{Introduction\label{introduction}}
Probing and manipulating quantum phase coherence are the heart of quantum information processes, and have long been studied for various mesoscopic systems\cite{imry}. One of the most powerful techniques to detect quantum phase coherence is to measure the phase difference using the Aharonov-Bohm (AB) effect\cite{ab,yacoby}. Recently the AB oscillations of the tunneling current through the quantum dot (QD) systems have been observed experimentally\cite{yacoby,hatano}. S\'{a}nchez \textit{et al}. have shown that the Coulomb interaction causes the magnetic field dependence in the transport properties of electrons which are not directly affected by a magnetic field\cite{sanchez}.

In this paper, we study the Coulomb interaction-induced AB oscillations in the linear conductance through a remote QD (RQD), which has no tunnel coupling with the QD embedded in an AB interferometer (ABI). Here ``Coulomb interaction-induced" means that electrons through the RQD do not acquire the AB phase directly and are affected only by capacitive coupling between the RQD and the QD embedded in the ABI. As a result, the transport properties through the RQD show the oscillations with respect to the magnetic flux threading through the ABI. Using an electronic Mach-Zehnder interferometer, such system had experimentally been realized\cite{mzi}. In particular, we investigate the visibility of Coulomb interaction-induced AB oscillations for weak and strong interaction regimes. In a weak interaction regime, we show that the Coulomb interaction-induced AB effect is characterized by a charge susceptibility of the RQD. In contrast, for a strong interaction regime, the Coulomb interaction-induced AB effect is not characterized by a charge susceptibility of the RQD due to many-body correlation effect. Moreover, we show that around the particle-hole symmetric point, there exists the region where the visibility of Coulomb interaction-induced AB oscillations is much higher than that of original AB oscillations in the ABI. At the infinitely strong interdot Coulomb interaction, when the two QD energy levels are equal, we discuss the QD energy level dependence and investigate the power-law behavior of visibility when the QD energy level is very far from the Fermi level.

The outline of this paper is as follows. In Sec. \ref{model}, we introduce a microscopic model Hamiltonian for an ABI containing a QD and a remote system with a RQD. Those two QDs are capacitively coupled, while no tunnel coupling exists. In Sec. \ref{formulation}, we provide the theoretical formulation to calculate the AB oscillations in the linear conductance through the RQD and its visibility. In particular, we employ the second-order perturbation theory in a weak interaction regime ($V_C\ll\hbar\Gamma$) and the decoupling approximatetion in the equation of motion approach for a strong interaction regime ($V_C\ge\hbar\Gamma$), respectively. Here $V_C$ and $\Gamma$ are the interdot Coulomb interaction strength and coupling strength between the QD and reservoirs, respectively. In Sec. \ref{induced}, we examine the Coulomb interaction-induced AB oscillations in the linear conductance through the RQD and the interdot Coulomb interaction dependences of the visibility both in weak and strong interaction regimes. Section \ref{summary} summarizes our results. In Appendix \ref{perturbation}, we calculate the retarded Green's functions of the QDs using the perturbation theory for weak interaction regime. In Appendix \ref{omega}, according to the decoupling scheme by Ref. \onlinecite{ora}, we estimate the self-energy by the higher-order correlation between the QD and the reservoir in the strong interaction regime. In Appendix \ref{population}, we discuss the phase of AB oscillations in the unperturbed population of the QD embedded in the ABI. We investigate the asymptotic behaviors of the visibility in $|\epsilon_0|\gg\hbar\Gamma$ at an infinitely large $V_C$ in Appendix \ref{asymptotic}. In Appendix \ref{ad-energy}, the QD energy dependence of the visibility near the Fermi level is shown.

\section{Model\label{model}}
\begin{figure}
\includegraphics[scale=0.4]{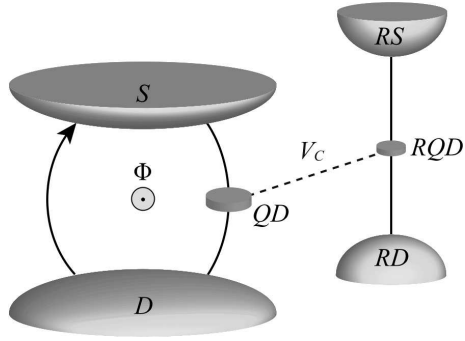}
\caption{\label{fig1} Schematic diagram of an Aharonov-Bohm interferometer containing a quantum dot (QD) which capacitively couples to a remote QD (RQD). $V_C$ is the strength of capacitive coupling between two QDs.}
\end{figure}

We consider an ABI containing a QD which capacitively couples to a RQD as shown in Fig. \ref{fig1}. We assume that the level spacing in QDs is much larger than other energy scales, and consider only a single energy level in each QD. To focus on the coherent charge transport, we neglect the spin degree of freedom. The Hamiltonian represents the sum of the following terms: $H=H_{ABI}+H_R+H_I$. The Hamiltonian $H_{ABI}$ describes the ABI given by
\begin{eqnarray}
H_{ABI}&=&\sum_{\nu\in\{S,D\}}\sum_k\epsilon_{\nu k}{a_{\nu k}}^{\dagger}a_{\nu k}+\epsilon_{AB}{c_{AB}}^{\dagger}c_{AB}\nonumber\\
&&+\sum_{\nu\in\{S,D\}}\sum_k\left(t_{\nu}{a_{\nu k}}^{\dagger}c_{AB}+\mbox{h.c.} \right)+\sum_{k,q}\left(|W|e^{i\phi}{a_{Sk}}^{\dagger}a_{Dq}+\mbox{h.c.} \right),
\end{eqnarray}
where $\epsilon_{\nu k}$ is the electron energy with wave number $k$ in the reservoir $\nu$, the operator $a_{\nu k}$ (${a_{\nu k}}^{\dagger}$) annihilates (creates) an electron in the reservoir $\nu$, $\epsilon_{AB}$ is the energy level of the QD embedded in the ABI, $c_{AB}$ (${c_{AB}}^{\dagger}$) is an annihilation (creation) operator of an electron in the QD, and $t_{\nu}$ is the tunneling amplitude between the QD and the reservoir $\nu$. The linewidth function of the QD level due to tunnel coupling to the reservoir $\nu$ is (excluding the effect of the direct tunneling $|W|$ between the source and drain reservoirs) $\Gamma_{AB}=\Gamma_S+\Gamma_D$ with $\Gamma_{S(D)}=(2\pi/\hbar)|t_{S(D)}|^2\rho_{S(D)}$, where $\rho_{S(D)}$ is the density of states in the reservoir $S$ $(D)$. Here we consider the wide-band limit and neglect the energy dependence of the linewidth function. In the last term, we define the magnetic flux dependent direct transmission between the source ($S$) and drain ($D$) reservoirs. Here we introduced the AB phase $\phi=2\pi\Phi/\Phi_0$, where $\Phi$ is the magnetic flux threading through the ABI and $\Phi_0=h/e$ is the magnetic flux quantum. The Hamiltonian $H_R$ represents the remote system including the RQD described by the non-interacting single impurity Anderson model,
\begin{equation}
H_R=\sum_{\nu_R\in\{RS,RD\}}\sum_{k_R}\epsilon_{\nu_Rk_R}{a_{\nu_Rk_R}}^{\dagger}a_{\nu_Rk_R}+\epsilon_d{c_d}^{\dagger}c_d+\sum_{\nu_R\in\{RS,RD\}}\sum_{k_R}\left(t_{\nu_R}{a_{\nu_Rk_R}}^{\dagger}c_d+\mbox{h.c.} \right),
\end{equation}
where $\epsilon_d$ is the energy level of the RQD, and $c_d$ (${c_d}^{\dagger}$) is an annihilation (creation) operator of an electron in the RQD, and $t_{\nu_R}$ is the tunneling amplitude between the RQD and the reservoir $\nu_R$. We introduce the linewidth function $\Gamma_d=\Gamma_{RS}+\Gamma_{RD}$. The interaction Hamiltonian is
\begin{equation}
H_I=V_Cn_{AB}n_d,
\end{equation}
where $V_C$ is the repulsive Coulomb interaction strength between the QD and the RQD, and $n_{AB}={c_{AB}}^{\dagger}c_{AB}$ and $n_d={c_d}^{\dagger}c_d$ are the number operators of QD and RQD, respectively.

\section{Formulation\label{formulation}}

\subsection{Transport through an ABI and remote system}
We consider the linear conductance through the ABI given by\cite{hofstetter}
\begin{eqnarray}
G_{AB}(\phi)&=&\frac{2e^2}{h}\int \frac{d\epsilon}{\hbar}\left[-\frac{\partial f(\epsilon)}{\partial\epsilon} \right]\left[\mathcal{T}_r+\frac{2}{1+x}\sqrt{\Gamma_S\Gamma_D\mathcal{T}_r(1-\mathcal{T}_r)}\cos\phi\cdot\mbox{Re}\{G_{AB}^r(\epsilon,\phi) \} \right.\nonumber\\
&&\left.-\frac{1}{2}\left\{\frac{4\Gamma_S\Gamma_D}{{\Gamma_{AB}}^2}(1-\mathcal{T}_r\cos^2\phi)-\mathcal{T}_r \right\}\tilde{\Gamma}_{AB}\cdot\mbox{Im}\{G_{AB}^r(\epsilon,\phi) \} \right].
\end{eqnarray}
Similarly, the linear conductance through the RQD is given by\cite{meir}
\begin{equation}
G_{RQD}(\phi)=\frac{2e^2}{h}\int d\epsilon\left[-\frac{\partial f(\epsilon)}{\partial\epsilon} \right]\frac{\Gamma_{RS}\Gamma_{RD}}{\Gamma_d}\left[-\mbox{Im}\left\{G_d^r(\epsilon,\phi) \right\} \right].\label{conductance}
\end{equation}
In the following, we choose the Fermi energy as the origin of energy. Here $f(\epsilon)=1/(e^{\epsilon/k_BT}+1)$ is an equilibrium Fermi-Dirac distribution function and $\tilde{\Gamma}_{AB}=\Gamma_{AB}/(1+x)$, where $x=\pi^2\rho_S\rho_D|W|^2$, and $\mathcal{T}_r=4x/(1+x)^2$ is the transmission probability for the direct transmission between the two reservoirs S and D. We assumed that the temperatures of four reservoirs are $T$. $G_{AB}^r(\epsilon,\phi)$ and $G_d^r(\epsilon,\phi)$ are the Fourier transform of the single-particle retarded Green's functions of the QD embedded in the ABI and the RQD, respectively,
\begin{eqnarray}
G_{AB}^r(t,t')&=&-i\theta(t-t')\left\langle \{c_{AB}(t),{c_{AB}}^{\dagger}(t') \} \right\rangle,\\
G_d^r(t,t')&=&-i\theta(t-t')\left\langle \{c_d(t),{c_d}^{\dagger}(t') \} \right\rangle.
\end{eqnarray}
As seen in the next section, the retarded Green's function of the RQD depends on the AB phase $\phi$ via the Coulomb interaction between the RQD and the QD embedded in the ABI and thus from Eq. (\ref{conductance}) the linear conductance through the RQD depends on the AB phase $\phi$. This is the origin of the Coulomb interaction-induced AB oscillations. The visibility of the oscillations in the linear conductance through the ABI (RQD) is defined as
\begin{equation}
\eta_{AB(RQD)}=\frac{\mbox{Max}\left[G_{AB(RQD)}(\phi)\right]-\mbox{Min}\left[G_{AB(RQD)}(\phi) \right]}{\mbox{Max}\left[G_{AB(RQD)}(\phi)\right]+\mbox{Min}\left[G_{AB(RQD)}(\phi) \right]}.\label{vis-def}
\end{equation}

\subsection{Green's functions}
Here we calculate the Green's function to estimate the transport properties discussed in the previous section.

\subsubsection{Weak interaction regime}
Here we consdier the weak interaction regime, namely $V_C\ll \hbar\Gamma_{d(AB)}$. We employ the perturbation theory with respect to $V_C$. Within the second-order perturbation theory, the single-particle retarded Green's function is given by
\begin{equation}
G_d^r(\epsilon,\phi)=g_d^r(\epsilon)+g_d^r(\epsilon)\Sigma_d^r(\epsilon,\phi)g_d^r(\epsilon),
\end{equation}
where the unperturbed retarded Green's function $g_d^r(\epsilon)$ and the retarded self-energy $\Sigma_d^r(\epsilon,\phi)$ are given in Appendix \ref{perturbation}.

\subsubsection{Strong interaction regime}
Here we consider the strong interaction regime, namely $V_C\gg\hbar\Gamma_{d(AB)}$. We employ the decoupling approximation in the equation of motion (EOM) for the retarded Green's function\cite{mwl,ora}.
\begin{equation}
i\hbar\frac{\partial}{\partial t}G_d^r(t,t')=\hbar\delta(t-t')+\epsilon_dG_d^r(t,t')+\sum_{\nu_R\in\{RS,RD \}}\sum_{k_R}{t_{\nu_R}}^*G_{\nu_Rk_R,d}^r(t,t')+V_CG_d^{r(2)}(t,t'),\label{eom1}
\end{equation}
where the two-particle retarded Green's function is defined as
\begin{eqnarray}
G_d^{r(2)}(t,t')&=&-i\theta(t-t')\left\langle \left\{c_d(t)n_{AB}(t),{c_d}^{\dagger}(t') \right\} \right\rangle,
\end{eqnarray}
and $G_{\nu_Rk_R,d}^r(t,t')$ is
\begin{equation}
G_{\nu_Rk_R,d}^r(t,t')=-i\theta(t-t')\left\langle \left\{a_{\nu_Rk_R}(t),{c_d}^{\dagger}(t') \right\} \right\rangle.
\end{equation}
From the EOM for $G_{\nu_Rk_R,d}^r(t,t')$, we obtain
\begin{equation}
G_{\nu_Rk_R,d}^r(t,t')=\int dt_1g_{\nu_Rk_R}^r(t,t_1)t_{\nu_R}G_d^r(t_1,t').
\end{equation}
Using the Fourier transformation, Eq. (\ref{eom1}) becomes
\begin{equation}
(\epsilon-\epsilon_d-\hbar\Sigma_d^r)G_d^r(\epsilon)=\hbar+V_CG_d^{r(2)}(\epsilon).
\end{equation}
Here the non-interacting tunneling retarded self-energy is given by
\begin{eqnarray}
\Sigma_d^{r(0)}&=&\sum_{\nu_R\in\{RS,RD \}}\sum_{k_R}|t_{\nu_R}|^2g_{\nu_Rk_R}^r(\epsilon)\nonumber\\
&=&-\frac{i}{2}\Gamma_d,
\end{eqnarray}
where $g_{\nu_Rk_R}^r(\epsilon)$ is the retarded Green's function of an isolated reservoir $\nu_R$. Similarly, we can calculate the EOM for $G_d^{r(2)}(t,t')$ as
\begin{eqnarray}
i\hbar\frac{\partial}{\partial t}G_d^{r(2)}(t,t')&=&\hbar\delta(t-t')\langle n_{AB} \rangle+(\epsilon_d+V_C)G_d^{r(2)}(t,t')+\sum_{\nu_R\in\{RS,RD\}}\sum_{k_R}{t_{\nu_R}}^*\Gamma_{1,\nu_Rk_R}^{(2)}(t,t')\nonumber\\
&&+\sum_{\nu\in\{S,D\}}\sum_k\left[t_{\nu}\Gamma_{2,\nu k}^{(2)}(t,t')-{t_{\nu}}^*\Gamma_{3,\nu k}^{(2)}(t,t') \right],
\end{eqnarray}
where the new retarded Green's functions are defined as
\begin{eqnarray}
\Gamma_{1,\nu_Rk_R}^{(2)}(t,t')&=&-i\theta(t-t')\left\langle \left\{a_{\nu_Rk_R}(t)n_{AB}(t),{c_d}^{\dagger}(t') \right\} \right\rangle,\\
\Gamma_{2,\nu k}^{(2)}(t,t')&=&-i\theta(t-t')\left\langle \left\{{a_{\nu k}}^{\dagger}(t)c_d(t)c_{AB}(t),{c_d}^{\dagger}(t') \right\} \right\rangle,\\
\Gamma_{3,\nu k}^{(2)}(t,t')&=&-i\theta(t-t')\left\langle \left\{a_{\nu k}(t){c_{AB}}^{\dagger}(t)c_d(t),{c_d}^{\dagger}(t') \right\} \right\rangle.
\end{eqnarray}
We use the following decoupling scheme by Ref. \onlinecite{mwl}
\begin{equation}
\Gamma_{1,\nu_Rk_R}^{(2)}(t,t')\simeq\langle n_{AB} \rangle G_{\nu_Rk_R,d}^r(t,t'),
\end{equation}
and $\Gamma_{2,\nu k}^{(2)}(t,t')=\Gamma_{3,\nu k}^{(2)}(t,t')=0$. Using the Fourier transformation, we have
\begin{equation}
(\epsilon-\epsilon_d-V_C)G_{d}^{r(2)}(\epsilon)=\hbar\langle n_{AB} \rangle+\hbar\langle n_{AB} \rangle\Sigma_d^{r(0)}G_d^r(\epsilon).\label{eom2}
\end{equation}
From Eqs. (\ref{eom1}) and (\ref{eom2}), $G_d^r(\epsilon)$ is given by
\begin{equation}
G_d^r(\epsilon)=\frac{\frac{\epsilon-\epsilon_d-V_C(1-\langle n_{AB}\rangle)}{\hbar}}{\frac{\epsilon-\epsilon_d}{\hbar}\frac{\epsilon-\epsilon_d-V_C}{\hbar}+\frac{i}{2}\Gamma_d\frac{\epsilon-\epsilon_d-V_C(1-\langle n_{AB} \rangle)}{\hbar}}.\label{grd}
\end{equation}
Similarly, we can calculate the retarded Green's function of the QD embedded in the ABI
\begin{equation}
G_{AB}^r(\epsilon)=\frac{\frac{\epsilon-\epsilon_{AB}-V_C(1-\langle n_d\rangle)}{\hbar}}{\frac{\epsilon-\epsilon_{AB}}{\hbar}\frac{\epsilon-\epsilon_{AB}-V_C}{\hbar}+\frac{1}{2}\left(\sqrt{\Gamma_S\Gamma_D\mathcal{T}_r}\cos\phi+i\tilde{\Gamma}_{AB} \right)\frac{\epsilon-\epsilon_{AB}-V_C(1-\langle n_d\rangle)}{\hbar}}.\label{gr-ab}
\end{equation}

These retarded Green's functions include the population of two QDs. In equilibrium, we can use the fluctuation-dissipation theorem,
\begin{equation}
\langle n_{d(AB)} \rangle=-\frac{1}{\pi}\int\frac{d\epsilon}{\hbar}f(\epsilon)\mbox{Im}\left\{G_{d(AB)}^r(\epsilon) \right\},\label{population-relation}
\end{equation}
to obtain a closed form for the population $\langle n_{d(AB)} \rangle$, and thus we can determine the retarded Green's functions. Using these results, we can calculate the linear conductances through the RQD and the ABI.

In Ref. \onlinecite{mwl}, the corraltions between the QD and the reservoir such as $\langle {a_{\nu_Rk_R}}^{\dagger}(t)c_d(t) \rangle$ had been disregarded. The decoupling scheme by Ref. \onlinecite{ora} takes account of those contributions. As a result, the retarded Green's function of the RQD is given by
\begin{equation}
G_d^r(\epsilon)=\frac{\frac{\epsilon-\epsilon_d-V_C(1-\langle n_{AB}\rangle)}{\hbar}}{\frac{\epsilon-\epsilon_d}{\hbar}\frac{\epsilon-\epsilon_d-V_C}{\hbar}+\frac{i}{2}\Gamma_d\frac{\epsilon-\epsilon_d-V_C(1-\langle n_{AB} \rangle)}{\hbar}-\frac{V_C}{\hbar}\Omega_d},
\end{equation}
where $\Omega_d$ is a pure imaginary additional energy given by
\begin{equation}
\Omega_d=-2i\sum_{\nu,k}t_{\nu}\mbox{Im}\left\{\langle {a_{\nu k}}^{\dagger}(t)c_{AB}(t) \rangle \right\}.\label{omega-def}
\end{equation}
Here $\langle {a_{\nu k}}^{\dagger}(t)c_{AB}(t) \rangle$ can be estimated by the fluctuation-dissipation theorem
\begin{equation}
\langle {a_{\nu k}}^{\dagger}(t)c_{AB}(t) \rangle=-\int\frac{d\epsilon}{2\pi i\hbar}f(\epsilon)\left\{G_{AB,\nu k}^r(\epsilon)-[G_{\nu k,AB}^r(\epsilon)]^* \right\}.
\end{equation}
However, as shown in Appendix \ref{omega}, we find that $\Omega_d=0$, in our model. Therefore, the two decoupling schemes by Refs. \onlinecite{mwl} and \onlinecite{ora} give the same results. Similarly, using the decoupling scheme by Ref. \onlinecite{ora}, the retarded Green's function of the QD embedded in the ABI is equivalent to Eq. (\ref{gr-ab}).

\section{Coulomb interaction-induced Aharonov-Bohm oscillations\label{induced}}
In the following, we focus on the situation when $\epsilon_{AB}=\epsilon_d\equiv\epsilon_0$, $\Gamma_{\nu}=\Gamma_{\nu_R}\equiv\Gamma/2$, $x=0.1$, and $T=0$. Here we discuss the induced AB oscillations in two regimes using the formulation in the previous section.

\subsection{Weak interaction regime}
\begin{figure}
\includegraphics[scale=0.45]{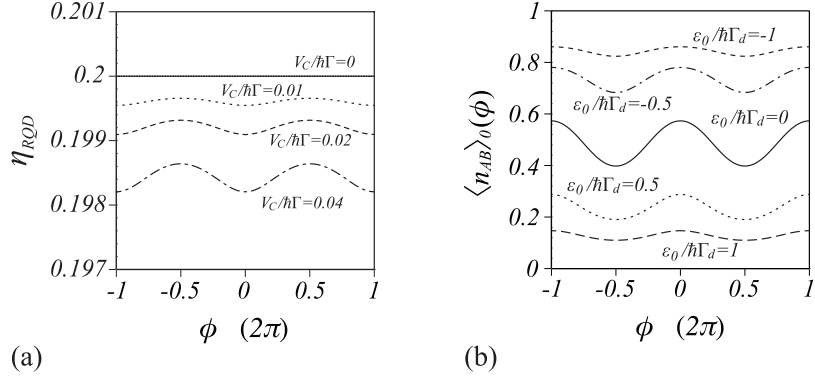}
\caption{\label{fig2} Induced AB oscillations and the unperturbed population of the QD embedded in the ABI. (a) Induced AB oscillations of $G_{RQD}(\phi)$ for various $V_C$ when $\epsilon_0=\hbar\Gamma$ and $x=0.1$. (b) AB oscillations in $\langle n_{AB}\rangle_0(\phi)$ for various QD energy levels.}
\end{figure}

We plot the interaction dependences of the induced AB oscillations in Fig. \ref{fig2}(a) when $\epsilon_0=0$. The period of oscillations is $2\pi$ and the linear conductance through the RQD is symmetric with respect to the AB phase since the linear conductance of a two-terminal system is an even function of the magnetic flux (AB phase), as required by Onsager-B\"{u}ttiker symmetry relations\cite{onsager,buttiker}.

To understand the origin of the induced AB oscillations, we consider the linear conductance through the RQD to the first-order of $V_C$ and $G_{RQD}(\phi)$ is given by
\begin{eqnarray}
G_{RQD}(\phi)&=&\frac{e^2}{h}\left(\frac{\Gamma}{2} \right)^2\left\{\frac{1}{\left(\frac{\epsilon_0}{\hbar} \right)^2+\left(\frac{\Gamma}{2} \right)^2}-\frac{2\frac{\epsilon_0}{\hbar}}{\left[\left(\frac{\epsilon_0}{\hbar} \right)^2+\left(\frac{\Gamma}{2} \right)^2 \right]^2}\frac{V_C}{\hbar}\langle n_{AB}\rangle_0(\phi)+O({V_C}^2) \right\}\nonumber\\
&\equiv&g_0(\epsilon_0)+\frac{\partial g_0(\epsilon_0)}{\partial\epsilon_0}\langle n_{AB}\rangle_0(\phi)\frac{V_C}{\hbar}+O({V_C}^2),\label{susceptibility}
\end{eqnarray}
where $g_0$ is the linear conductance through the RQD without $V_C$ given by
\begin{equation}
g_0(\epsilon_0)\equiv\frac{e^2}{h}\frac{\left(\frac{\Gamma}{2} \right)^2}{\left(\frac{\epsilon_0}{\hbar} \right)^2+\left(\frac{\Gamma}{2} \right)^2},
\end{equation}
and the AB flux dependence of the linear conductance through the RQD only appears in an unperturbed population $\langle n_{AB}\rangle_0(\phi)$ defined in Eq. (\ref{nab0}).  As shown in Fig. \ref{fig2}(b), $\langle n_{AB}\rangle_0(\phi)$ oscillates with the flux $\phi$. The second term of right-hand side in Eq. (\ref{susceptibility}) shows that the Coulomb interaction-induced AB oscillations in the linear conductance through the RQD is characterized by a charge susceptibility of the RQD $\partial g_0/\partial\epsilon_0$ which is the change of the conductance by the change of energy level of the RQD induced by the charge in the QD embedded in the ABI.

From Eq. (\ref{susceptibility}), for $\epsilon_0=0$, we find that the first-order contribution is absent. For $\epsilon_0\neq 0$, using the right-hand side in the first line of Eq. (\ref{susceptibility}), the visibility for the induced AB oscillations is expressed as
\begin{equation}
\eta_{RQD}=\mbox{sign}(\epsilon_0)\frac{G_{RQD}(\phi=\pi)-G_{RQD}(\phi=0)}{G_{RQD}(\phi=\pi)+G_{RQD}(\phi=0)},\label{visibility-def}
\end{equation}
since the unperturbed population $\langle n_{AB}\rangle_0(\phi)$ has a peak at $\phi=0$ as proven in Appendix \ref{population} (see Fig. \ref{fig2}(b)).

\begin{figure}
\includegraphics[scale=0.45]{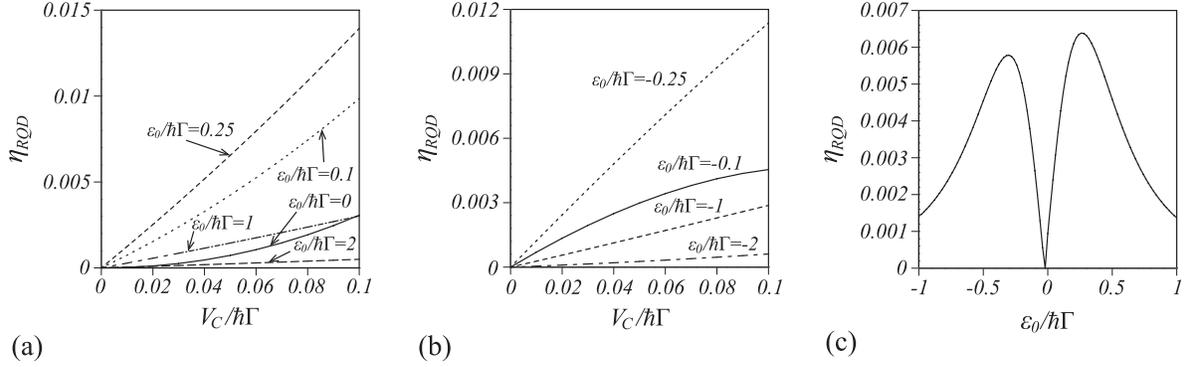}
\caption{\label{fig3} Interaction dependences of the visibility for the Coulomb interaction-induced AB oscillations in the linear conductance through the RQD for various energy levels $\epsilon_{AB}=\epsilon_d=\epsilon_0$ when $\Gamma_{\nu}=\Gamma/2$ and $x=0.1$. (a) $\epsilon_0\ge0$. (b) $\epsilon_0<0$. (c) $\epsilon_0$ dependence of the visibility for $V_C/\hbar\Gamma=0.05$.}
\end{figure}

Within the second-order perturbation theory, we plot the interaction dependence of the visibility of the induced AB oscillations for various values of energy level, $\epsilon_0$, in Figs. \ref{fig3}(a) and (b). For $\epsilon_0\neq 0$, the visibility increases linearly when the interaction strength increases for $V_C\ll\hbar\Gamma$. Furthermore, the visibility decreases when the energy level goes away from the Fermi level ($|\epsilon_0|\gtrsim0.3\hbar\Gamma$). This can be seen from $\epsilon_0$ dependence of the visibility for a fixed $V_C$ as shown in Fig. \ref{fig3}(c). For $\epsilon_0=0$, the first-order contribution vanishes and thus the visibility increases parabolically with respect to $V_C$. We find that the visibility increases with $|\epsilon_0|$ when the energy level is close to the Fermi level ($\epsilon_0=0$).

In the previous study\cite{kubo}, using the nonequilibrium second-order perturbation theory for $V_C$, we investigated the backaction dephasing by the QD detector. The backaction dephasing rate is defined as the imaginary part of the retarded self-energy given in Eq. (\ref{self-d}). In Ref. \onlinecite{kubo}, we clarified that the origin of the backaction by the QD detector is its charge noise. Unlike the formulation of present paper, we had compensated the energy level shift by the real part of the self-energy to discuss only the backaction dephasing. In the previous study, we had discussed the visibility of AB oscillations in the linear conductance in the measured system (ABI). In this paper, in contrast, we focus on the visibility of oscillations in the linear conductance through the RQD corresponding to the QD detector.

\subsection{Strong interaction regime}
\begin{figure}
\includegraphics[scale=0.5]{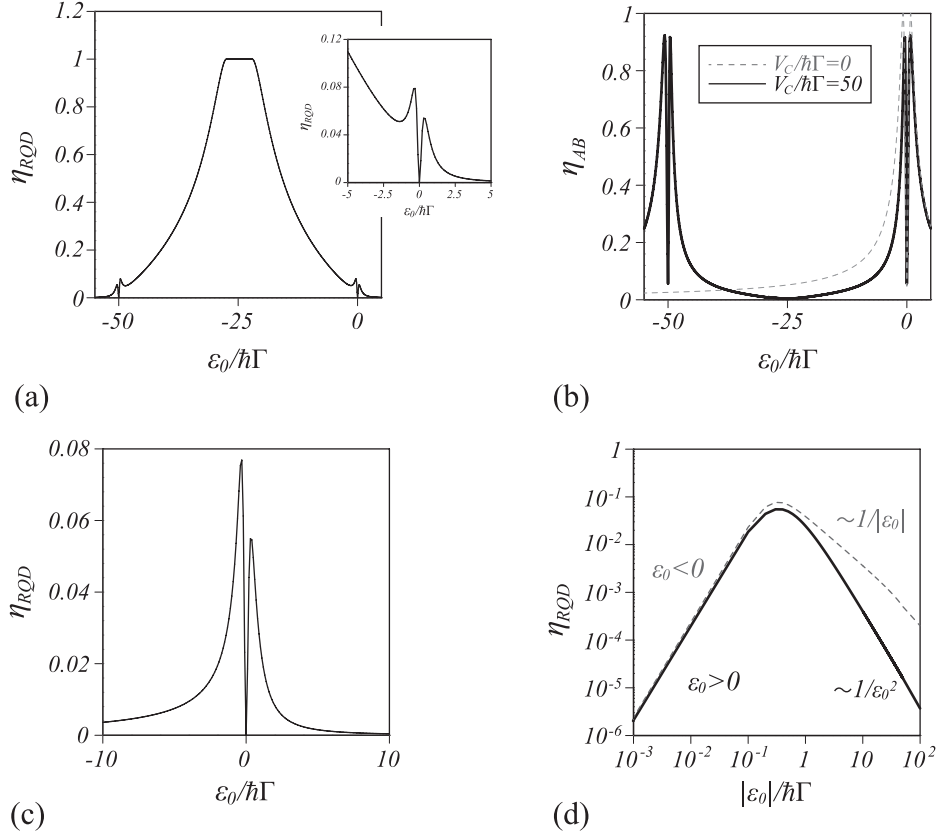}
\caption{\label{fig4} QD energy dependences of the visibility for the Coulomb interaction-induced AB oscillations in the linear conductance through the RQD when $\epsilon_{AB}=\epsilon_d=\epsilon_0$, $\Gamma_{\nu}=\Gamma/2$, and $x=0.1$. (a) In the finite Coulomb interaction ($V_C/\hbar\Gamma=50$). (b) Visibility $\eta'$ of AB oscillations in the linear conductance through the ABI without $V_C$. (c) In the strong Coulomb interaction limit ($V_C\to\infty$), Inset: Near the Fermi level. (d) Log-log plot of (c).}
\end{figure}
In Fig. \ref{fig4}(a), we plot the numerical results for the QD energy dependence of the visibility when $V_C/\hbar\Gamma=50$. At $\epsilon_0=0$ and $\epsilon_0=-V_C$, the visibility vanishes since the linear conductance through the RQD is $G_{RQD}=e^2/h$ which is independent of the AB phase from Eqs. (\ref{conductance}) and (\ref{grd}). This result at zero temperature is very special and the visibility at $\epsilon_0$ and $\epsilon_0=-V_C$ is non-zero in finite temperatures. Surprisingly, around the particle-hole symmetric point $\epsilon_0=-V_C/2$, the visibility of remote system becomes $1$ although the visibility of original AB oscillation in the ABI is quite low as shown in Fig. \ref{fig4}(b). The visibility becomes $1$ when the minimum value of $G_{RQD}$ is equal to zero, namely from Eqs. (\ref{conductance}) and (\ref{grd}), we find that $G_{RQD}=0$ when $\epsilon_0=-V_C(1-\langle n_{AB}\rangle)$. Near the particle-hole symmetric point $\epsilon_0=-V_C/2$, we have $\langle n_{AB}\rangle\simeq 1/2$. As a result, for $\epsilon_0\simeq-V_C/2$, the visibility reaches $1$. Without Coulomb interaction, the visibility of AB oscillation in the ABI has a double peak near the Fermi level ($\epsilon_0=0$) as shown in Fig. 4(b) (thin dashed-line). It is well-known that the transmission probability through the ABI can be $0$ due to the Fano anti-resonance\cite{hofstetter,fano,kobayashi}. As a result, the conductance through the ABI becomes zero at this resonance. Thus, the peak height of the double peak in the visibility is $1$. In contrast, for finite Coulomb interaction ($V_C/\hbar\Gamma=50$), the peak height of the double peak is less than $1$ and the visibility decreases because of the Coulomb interaction effect (solid line in Fig. 4(b)). In Fig. \ref{fig4}(c), we show the QD energy level dependences of the visibility when $V_C$ is infinitely large. When $|\epsilon_0|$ is much larger than $\Gamma$, the visibility shows the power-law behavior of ${\epsilon_0}^{-2}$ for $\epsilon_0>0$ and $|\epsilon_0|^{-1}$ for $\epsilon_0<0$. In Fig. \ref{fig4} (d), we show the log-log plot of Fig. \ref{fig4} (c). We find that the slope of $\eta$ for $\epsilon_0>0$ is $-2$ and that of $\eta$ for $\epsilon_0<0$ is $-1$ in the region of $|\epsilon_0|\gg\hbar\Gamma$. The visibility has two peaks as a function of QD energy level $\epsilon_0$ since the visibility is zero at $\epsilon_0=0$.  Near $\epsilon_0=0$, the visibility shows the power-law behavior of ${\epsilon_0}^2$ as discussed in Appendix \ref{ad-energy}. This behavior is clearly different from the linear dependence for $\epsilon_0$ near the Fermi level characterized by the charge susceptibility found in the weak interaction regime. Thus, in the strong correlation limit, the Coulomb interaction-induced AB oscillation does not relate with the charge susceptibility of the RQD.

Even in the strong interaction limit, we have the finite visibility of AB oscillations in the linear conductance through the ABI. Although the QD in the ABI strongly couples to the RQD which could play a role of the charge detector, the coherence in the ABI remains finite since the detector resolution of RQD is very low at very low source-drain bias voltage ($V_{SD}\simeq0$, namely linear response regime) and the RQD cannot accurately measure the charge of QD in the ABI. As a result, quantum interference effect remains since we cannot determine which path the electron goes through.

\section{Summary\label{summary}}
To summarize, we have studied the Coulomb interaction-induced AB oscillations in the transport through a RQD which is capacitively coupled to the QD embedded in an ABI. In particular, in a weak interaction regime, we have shown that the charge susceptibility of the RQD characterizes the Coulomb interaction-induced AB oscillations. The visibility increases linearly with respect to the interdot Coulomb interaction except when the QD energy level align the Fermi level ($\epsilon_0=0$). For $\epsilon_0=0$, the visibility shows the parabolic dependence on $V_C$. In a strong but finite interaction regime, around the particle-hole symmetric point, there exists the region where the visibility of Coulomb interaction-induced AB oscillation is much higher than that of original AB oscillations in the ABI. In the strong interaction limit, when $\epsilon_0\gg\hbar\Gamma$, the visibility shows the power-law behavior of ${\epsilon_0}^{-2}$. While for sufficiently negative $\epsilon_0$, the visibility shows the power-law behavior of ${|\epsilon_0|}^{-1}$. Moreover, the visibility has two peaks as a function of QD energy level $\epsilon_0$ since the visibility is zero at $\epsilon_0=0$. Near the Fermi level, the visibility shows the power-law behavior of ${\epsilon_0}^2$.

\begin{acknowledgements}
We thank W. G. van der Wiel and F. Morikoshi for useful comments and fruitful discussions. Part of this work is supported financially by JSPS MEXT Grant-in-Aid for Scientific Research on Innovative Areas (21102003), Funding Program for World-Leading Innovative R\&D Science and Technology (FIRST), and JSPS KAKENHI (26247051, 26870080).
\end{acknowledgements}

\appendix
\section{Retarded Green's functions in weak interaction regime\label{perturbation}}
\begin{figure}
\includegraphics[scale=0.5]{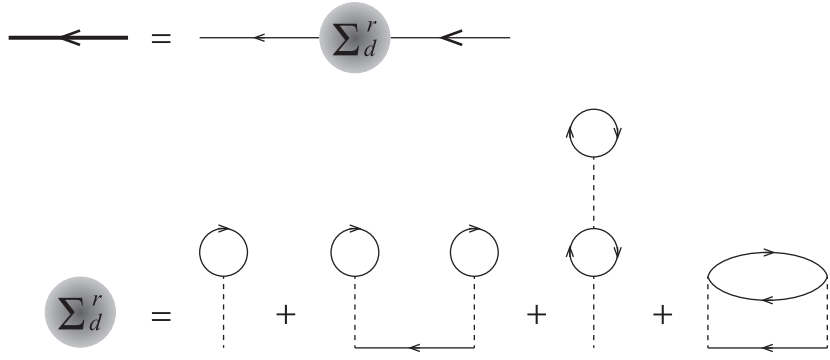}
\caption{\label{fig5} Feynman diagram for the Dyson's equation and the second-order self-energy with respect to the Coulomb interaction $V_C$. The solid, bold, and dashed lines correspond to the unperturbed, full Green's functions, and the Coulomb interaction.}
\end{figure}

Within the second-order perturbation theory, the Feynman diagram for the retarded self-energy is shown Fig. \ref{fig5}, and its expression is given by
\begin{eqnarray}
\Sigma_d^r(\epsilon,\phi)&=&\frac{V_C}{\hbar}\langle n_{AB} \rangle_0(\phi)\nonumber\\
&&+\left(\frac{V_C}{\hbar} \right)^2\int\frac{dE_1}{2\pi\hbar}\int\frac{dE_2}{2\pi\hbar}[g_{AB}^{--}(E_1,\phi)]^2g_d^{--}(E_2)\nonumber\\
&&-\left(\frac{V_C}{\hbar} \right)^2\int\frac{dE_1}{2\pi\hbar}\int\frac{dE_2}{2\pi\hbar}g_{AB}^{-+}(E_1,\phi)g_{AB}^{+-}(E_1,\phi)g_d^{++}(E_2)\nonumber\\
&&+\left(\frac{V_C}{\hbar} \right)^2\int\frac{dE_1}{2\pi\hbar}\int\frac{dE_2}{2\pi\hbar}\left[g_d^r(E_1)g_{AB}^{+-}(E_2,\phi)g_{AB}^{-+}(E_1+E_2-\epsilon,\phi) \right.\nonumber\\
&&+g_d^{-+}(E_1)g_{AB}^r(E_2,\phi)g_{AB}^{+-}(E_1+E_2-\epsilon,\phi)\nonumber\\
&&\left. +g_d^{-+}(E_1)g_{AB}^{+-}(E_2,\phi)g_{AB}^a(E_1+E_2-\epsilon,\phi)\right],\label{self-d}
\end{eqnarray}
where the unperturbed Green's functions are given by
\begin{eqnarray}
g_d^r(\epsilon)&=&\frac{1}{\frac{\epsilon-\epsilon_d}{\hbar}+\frac{i}{2}\Gamma_d}=[g_d^a(\epsilon)]^*,\\
g_d^{-+}(\epsilon)&=&-2if(\epsilon)\mbox{Im}\{g_d^r(\epsilon) \},\\
g_d^{+-}(\epsilon)&=&2i[1-f(\epsilon)]\mbox{Im}\{g_d^r(\epsilon) \},\\
g_d^{--}(\epsilon)&=&g_d^r(\epsilon)+g_d^{-+}(\epsilon),\\
g_{AB}^r(\epsilon,\phi)&=&\frac{1}{\frac{\epsilon-\epsilon_{AB}}{\hbar}+\frac{1}{2}\sqrt{\Gamma_S\Gamma_D\mathcal{T}_r}\cos\phi+\frac{i}{2}\tilde{\Gamma}_{AB}}=[g_{AB}^a(\epsilon,\phi)]^*,\label{grab0}\\
g_{AB}^{-+}(\epsilon,\phi)&=&-2if(\epsilon)\mbox{Im}\{g_{AB}^r(\epsilon,\phi) \},\\
g_{AB}^{+-}(\epsilon,\phi)&=&2i[1-f(\epsilon)]\mbox{Im}\{g_{AB}^r(\epsilon,\phi) \},\\
g_{AB}^{--}(\epsilon,\phi)&=&g_{AB}^r(\epsilon,\phi)+g_{AB}^{-+}(\epsilon,\phi),
\end{eqnarray}
and the unperturbed population is
\begin{equation}
\langle n_{AB}\rangle_0=-\frac{1}{\pi}\int\frac{d\epsilon}{\hbar}f(\epsilon)\mbox{Im}\{g_{AB}^r(\epsilon,\phi) \}.\label{nab0}
\end{equation}

Similarly, the retarded self-energy $\Sigma_{AB}^r(\epsilon,\phi)$ is given by
\begin{eqnarray}
\Sigma_{AB}^r(\epsilon,\phi)&=&\frac{V_C}{\hbar}\langle n_d \rangle_0(\phi)\nonumber\\
&&+\left(\frac{V_C}{\hbar} \right)^2\int\frac{dE_1}{2\pi\hbar}\int\frac{dE_2}{2\pi\hbar}[g_d^{--}(E_1,\phi)]^2g_{AB}^{--}(E_2)\nonumber\\
&&-\left(\frac{V_C}{\hbar} \right)^2\int\frac{dE_1}{2\pi\hbar}\int\frac{dE_2}{2\pi\hbar}g_d^{-+}(E_1,\phi)g_d^{+-}(E_1,\phi)g_{AB}^{++}(E_2)\nonumber\\
&&+\left(\frac{V_C}{\hbar} \right)^2\int\frac{dE_1}{2\pi\hbar}\int\frac{dE_2}{2\pi\hbar}\left[g_{AB}^r(E_1)g_d^{+-}(E_2,\phi)g_d^{-+}(E_1+E_2-\epsilon,\phi) \right.\nonumber\\
&&+g_{AB}^{-+}(E_1)g_d^r(E_2,\phi)g_d^{+-}(E_1+E_2-\epsilon,\phi)\nonumber\\
&&\left. +g_{AB}^{-+}(E_1)g_d^{+-}(E_2,\phi)g_d^a(E_1+E_2-\epsilon,\phi)\right],\nonumber\\
&&
\end{eqnarray}
and the unperturbed population is
\begin{equation}
\langle n_d\rangle_0=-\frac{1}{\pi}\int\frac{d\epsilon}{\hbar}f(\epsilon)\mbox{Im}\{g_d^r(\epsilon) \}.
\end{equation}

\section{Calculation of additional self-energy $\Omega_d$\label{omega}}
To evaluate the additional self-energy $\Omega_d$ defined as
\begin{equation}
\Omega_d=-2i\sum_{\nu,k}t_{\nu}\mbox{Im}\left\{\left\langle {a_{\nu k}}^{\dagger}(t)c_{AB}(t) \right\rangle \right\},
\end{equation}
using the fluctuation-dissipation theorem, we have to calculate two kinds of retarded Green's functions, $G_{AB,\nu k}^r(t,t')$ and $G_{\nu k,AB}^r(t,t')$. 

First we consider the EOM for the retarded Green's function $G_{AB,\nu k}^r(t,t')=-i\theta(t-t')\left\langle \{c_{AB}(t),{a_{\nu k}}^{\dagger}(t') \} \right\rangle$,
\begin{equation}
i\hbar\frac{\partial}{\partial t}G_{AB,\nu k}^r(t,t')=\epsilon_{AB}G_{AB,\nu k}^r(t,t')+\sum_{\nu'\in\{S,D \}}\sum_{k'}t_{\nu'}G_{\nu'k',\nu k}^r(t,t')+V_CG_{AB,\nu k}^{r(2)}(t,t'),
\end{equation}
where
\begin{eqnarray}
G_{\nu'k',\nu k}^r(t,t')&=&-i\theta(t-t')\left\langle \{a_{\nu'k'}(t),{a_{\nu k}}^{\dagger}(t') \} \right\rangle,\\
G_{AB,\nu k}^{r(2)}(t,t')&=&-i\theta(t-t')\left\langle \{c_{AB}(t)n_d(t),{a_{\nu k}}^{\dagger}(t') \} \right\rangle.
\end{eqnarray}
Using the same decoupling scheme as Ref. \onlinecite{ora}, we obtain
\begin{equation}
G_{AB,\nu k}^{r(2)}(t,t')\simeq\left\langle n_d\right\rangle G_{AB,\nu k}^r(t,t').
\end{equation}
After Fourier transform, the EOM of $G_{Ab,\nu k}^r(t,t')$ is
\begin{equation}
(\epsilon-\epsilon_{AB}-V_C\langle n_d\rangle)G_{AB,\nu k}^r(\epsilon)=\sum_{\nu'\in\{S,D \}}\sum_{k'}t_{\nu'}G_{\nu'k',\nu k}^r(\epsilon).\label{eom-a1}
\end{equation}
To estimate this, we consider the EOM of $G_{\nu'k',\nu k}^r(t,t')$. After Fourier transform, we obtain
\begin{eqnarray}
G_{\nu'k',\nu k}^r(\epsilon)&=&\delta_{\nu,\nu'}\delta_{k,k'}g_{\nu k}^r(\epsilon)+t_{\nu'}g_{\nu'k'}^r(\epsilon)G_{AB,\nu k}^r(\epsilon)\nonumber\\
&&+\delta_{\nu',S}\sum_{p,q}\delta_{k',p}|W|e^{i\phi}g_{\nu'k'}^r(\epsilon)G_{Dq,\nu k}^r(\epsilon)\nonumber\\
&&+\delta_{\nu',D}\sum_{p,q}\delta_{k',q}|W|e^{-i\phi}g_{\nu'k'}^r(\epsilon)G_{Sq,\nu k}^r(\epsilon).
\end{eqnarray}
Thus, we have
\begin{eqnarray}
  \left(\begin{array}{c}
    \sum_{k'}G_{Sk',\nu k}^r(\epsilon) \\ 
    \sum_{k'}G_{Dk',\nu k}^r(\epsilon) \\ 
  \end{array}\right)&=&\frac{1}{1+x}\left(  \begin{array}{c}
    \delta_{\nu,S}g_{Sk}^r(\epsilon)-i\pi\rho_S|W|e^{i\phi}\delta_{\nu,D}g_{Dk}^r(\epsilon) \\ 
    \delta_{\nu,D}g_{Dk}^r(\epsilon)-i\pi\rho_D|W|e^{-i\phi}\delta_{\nu,S}g_{Sk}^r(\epsilon) \\ 
  \end{array} \right)\nonumber\\
  &&+\frac{1}{1+x}\left(  \begin{array}{c}
    -i\pi\rho_St_S-\pi^2\rho_S\rho_Dt_D|W|e^{i\phi} \\ 
    -i\pi\rho_Dt_D-\pi^2\rho_S\rho_Dt_S|W|e^{-i\phi} \\ 
  \end{array} \right)G_{AB,\nu k}^r(\epsilon).
\end{eqnarray}
Here we use the relation
\begin{equation}
\sum_kg_{\nu k}^r(\epsilon)=-i\pi\rho_{\nu}.
\end{equation}
Therefore, Eq. (\ref{eom-a1}) is
\begin{eqnarray}
(\epsilon-\epsilon_{AB}-V_C\langle n_d\rangle)G_{AB,\nu k}^r(\epsilon)&=&\frac{1}{1+x}\left[\delta_{\nu,S}t_Sg_{Sk}^r(\epsilon)+\delta_{\nu,D}t_Dg_{Dk}^r(\epsilon) \right.\nonumber\\
&&\left. -i\pi\rho_St_S|W|e^{i\phi}\delta_{\nu,D}g_{Dk}^r(\epsilon)-i\pi\rho_Dt_D|W|e^{-i\phi}\delta_{\nu,S}g_{Sk}^r(\epsilon) \right]\nonumber\\
&&+\hbar\Sigma_{AB}^{r(0)}G_{AB,\nu k}^r(\epsilon),
\end{eqnarray}
where
\begin{equation}
\Sigma_{AB}^{r(0)}=-\frac{i}{2}\tilde{\Gamma}_{AB}-\frac{1}{2}\sqrt{\Gamma_S\Gamma_D\mathcal{T}_r}\cos\phi.
\end{equation}
Finally we obtain
\begin{equation}
\sum_{\nu\in\{S,D \}}\sum_kt_{\nu}G_{AB,\nu k}^r(\epsilon)=\frac{\Sigma_{AB}^{r(0)}}{\frac{\epsilon-\epsilon_{AB}-V_C\langle n_d\rangle}{\hbar}-\Sigma_{AB}^{r(0)}}.
\end{equation}

Similarly, from the EOM for $G_{\nu k,AB}^r(\epsilon)$, we obtain
\begin{equation}
\sum_{\nu\in\{S,D \}}\sum_kt_{\nu}\left[G_{\nu k,AB}^r(\epsilon) \right]^*=\frac{\left[\Sigma_{AB}^{r(0)} \right]^*}{\frac{\epsilon-\epsilon_{AB}-V_C\langle n_d\rangle}{\hbar}-\left[\Sigma_{AB}^{r(0)} \right]^*}.
\end{equation}

Using the above results, the additional self-energy $\Omega_d$ is
\begin{eqnarray}
\Omega_d&=&2i\sum_{\nu\in\{S,D \}}\sum_k\mbox{Im}\left\{\int\frac{d\epsilon}{2\pi i\hbar}f(\epsilon)\left(G_{AB,\nu k}^r(\epsilon)-\left[G_{\nu k,AB}^r(\epsilon) \right]^* \right) \right\}\nonumber\\
&=&-\frac{i}{\pi}\int\frac{d\epsilon}{\hbar}f(\epsilon)\mbox{Re}\left\{\frac{\Sigma_{AB}^{r(0)}}{\frac{\epsilon-\epsilon_{AB}-V_C\langle n_d\rangle}{\hbar}-\Sigma_{AB}^{r(0)}}-\frac{\left[\Sigma_{AB}^{r(0)} \right]^*}{\frac{\epsilon-\epsilon_{AB}-V_C\langle n_d\rangle}{\hbar}-\left[\Sigma_{AB}^{r(0)} \right]^*} \right\}\nonumber\\
&=&0.
\end{eqnarray}

\section{Unperturbed Population $\langle n_{AB}\rangle_0(\phi)$\label{population}}
In this Appendix, we discuss the unperturbed population $\langle n_{AB}\rangle_0(\phi)$.

Here we estimate the sign of the following quantity to determine the phase of AB oscillations of the unperturbed population at $\phi=0$, from Eqs. (\ref{grab0}) and (\ref{nab0})
\begin{eqnarray}
\left. \frac{\partial^2\langle n_{AB}\rangle_0(\phi)}{\partial\phi^2}\right|_{\phi=0}&=&\frac{\tilde{\Gamma}_{AB}}{2\pi}\sqrt{\Gamma_S\Gamma_D\mathcal{T}_r}\int_{-\infty}^{\infty}\frac{d\epsilon}{\hbar}f(\epsilon)\frac{\frac{\epsilon-\epsilon_{AB}}{\hbar}+\frac{1}{2}\sqrt{\Gamma_D\Gamma_D\mathcal{T}_r}}{\left[\left(\frac{\epsilon-\epsilon_{AB}}{\hbar}+\frac{1}{2}\sqrt{\Gamma_S\Gamma_D\mathcal{T}_r} \right)^2+\left(\frac{\tilde{\Gamma}_{AB}}{2} \right)^2 \right]^2}\nonumber\\
&=&-\frac{\hbar\tilde{\Gamma}_{AB}}{4\pi(k_BT)^2}\sqrt{\Gamma_S\Gamma_D\mathcal{T}_r}\int_{-\infty}^{\infty}\frac{d\epsilon}{k_BT}\frac{e^{\epsilon/k_BT}}{(e^{\epsilon/k_BT}+1)^2}\nonumber\\
&&\times\frac{1}{\left(\frac{\epsilon-\epsilon_{AB}+\frac{1}{2}\hbar\sqrt{\Gamma_S\Gamma_D\mathcal{T}_r}}{k_BT} \right)^2+\left(\frac{\hbar\tilde{\Gamma}_{AB}}{2k_BT} \right)^2}.\label{derivative}
\end{eqnarray}
Here we used partial integration. The sign of the right-hand side of Eq. (\ref{derivative}) is always negative since the integrand is positive definite. As a result, the unperturbed population $\langle n_{AB}\rangle_0(\phi)$ has a peak at $\phi=0$.

\section{Asymptotic behaviors of visibility in $|\epsilon_0|\gg\hbar\Gamma$\label{asymptotic}}
Here we estimate the visibility at an infinitely large inter-dot interaction $V_C$. From Eqs. (\ref{grd}) and (\ref{gr-ab}), we have the retarded Green's functions of two QDs at the infinitely large $V_C$
\begin{eqnarray}
G_d^r(\epsilon)&=&\frac{1-\langle n_{AB} \rangle}{\frac{\epsilon-\epsilon_0}{\hbar}+\frac{i}{2}\Gamma_d(1-\langle n_{AB}\rangle)}\\
G_{AB}^r(\epsilon)&=&\frac{1-\langle n_d\rangle}{\frac{\epsilon-\epsilon_0}{\hbar}+\frac{1}{2}(\gamma_a\cos\phi+i\tilde{\Gamma}_{AB})(1-\langle n_d\rangle)},
\end{eqnarray}
where $\gamma_a\equiv\sqrt{\Gamma_S\Gamma_D\mathcal{T}_r}$. Using the fluctuation-dissipation theorem, the average populations of two QDs are derived
\begin{eqnarray}
\langle n_d\rangle&=&\frac{1}{\pi}\int_{-\infty}^0\frac{d\epsilon}{\hbar}\frac{\frac{\Gamma_d}{2}(1-\langle n_{AB}\rangle)^2}{\left(\frac{\epsilon-\epsilon_0}{\hbar} \right)^2+\left[\frac{\Gamma_d}{2}(1-\langle n_{AB}\rangle) \right]^2},\\
\langle n_{AB}\rangle&=&\frac{1}{\pi}\int_{-\infty}^0\frac{d\epsilon}{\hbar}\frac{\frac{\tilde{\Gamma}_{AB}}{2}(1-\langle n_d\rangle)^2}{\left[\frac{\epsilon-\epsilon_0}{\hbar}+\frac{\gamma_a}{2}\cos\phi(1-\langle n_d\rangle) \right]^2+\left[\frac{\tilde{\Gamma}_{AB}}{2}(1-\langle n_d\rangle) \right]^2}.
\end{eqnarray}
The linear conductance through the RQD is
\begin{equation}
G_{RQD}(\phi)=\frac{e^2}{h}\frac{\left[\frac{\Gamma_d}{2}(1-\langle n_{AB}\rangle) \right]^2}{\left(\frac{\epsilon_0}{\hbar} \right)^2+\left[\frac{\Gamma_d}{2}(1-\langle n_{AB}\rangle) \right]^2}.\label{ap-cond}
\end{equation}

We assume the following form of $\langle n_{AB}\rangle$ as
\begin{equation}
\langle n_{AB}\rangle\sim\langle n_{AB}\rangle_{\pi/2}+\delta\cos\phi,
\end{equation}
where $\langle n_{AB}\rangle_{\pi/2}$ is the population when $\phi=\pi/2$. We also claim $|\delta|\ll\langle n_{AB}\rangle_{\pi/2}$, $1-\langle n_{AB}\rangle_{\pi/2}$, which should be checked in the following arguments. Putting this into Eq. (\ref{ap-cond}), we obtain
\begin{eqnarray}
G_{RQD}(\phi)&\sim&G_{RQD}\left(\frac{\pi}{2} \right)(1-A\delta\cos\phi),
\end{eqnarray}
where
\begin{eqnarray}
A&\equiv&\frac{2\left(\frac{2\epsilon_0}{\hbar\Gamma_d} \right)^2}{(1-\langle n_{AB}\rangle_{\pi/2})\left[\left(\frac{2\epsilon_0}{\hbar\Gamma_d} \right)^2+(1-\langle n_{AB}\rangle_{\pi/2})^2 \right]}>0.
\end{eqnarray}
From the definition of the visibility (\ref{vis-def}), we have
\begin{eqnarray}
\eta&=&\frac{2|\delta|\left(\frac{2\epsilon_0}{\hbar\Gamma_d} \right)^2}{(1-\langle n_{AB}\rangle_{\pi/2})\left[\left(\frac{2\epsilon_0}{\hbar\Gamma_d} \right)^2+(1-\langle n_{AB}\rangle_{\pi/2})^2 \right]}.
\end{eqnarray}
Clearly, the visibility is zero for $\epsilon_0=0$. For further discussions, we will evaluate $\delta$ and $\langle n_{AB}\rangle_{\pi/2}$, by solving following coupled equations:
\begin{eqnarray}
\langle n_{AB}\rangle_{\pi/2}&=&\frac{1}{\pi}\int_{-\infty}^0\frac{d\epsilon}{\hbar}\frac{\frac{\tilde{\Gamma}_{AB}}{2}(1-\langle n_d\rangle_{\pi/2})^2}{\left(\frac{\epsilon-\epsilon_0}{\hbar} \right)^2+\left[\frac{\tilde{\Gamma}_{AB}}{2}(1-\langle n_d\rangle_{\pi/2}) \right]^2},\\
\langle n_d\rangle_{\pi/2}&=&\frac{1}{\pi}\int_{-\infty}^0\frac{d\epsilon}{\hbar}\frac{\frac{\Gamma_d}{2}(1-\langle n_{AB}\rangle_{\pi/2})^2}{\left(\frac{\epsilon-\epsilon_0}{\hbar} \right)^2+\left[\frac{\Gamma_d}{2}(1-\langle n_{AB}\rangle_{\pi/2}) \right]^2}.
\end{eqnarray}
We restrict ourselves to the energy levels far from the Fermi energy, namely $|\epsilon_0|\gg\tilde{\Gamma}_{AB}/2$, $\Gamma_d/2$. First we consider the situation of $\epsilon_0>0$. The following definite integral for positive $\epsilon_0\gg\gamma_0$ is
\begin{eqnarray}
I[\gamma_0]\equiv\frac{1}{\pi}\int_{-\infty}^0\frac{d\epsilon}{\hbar}\frac{\gamma_0}{\left(\frac{\epsilon-\epsilon_0}{\hbar} \right)^2+{\gamma_0}^2}\sim\frac{\hbar\gamma_0}{\pi\epsilon_0}.
\end{eqnarray}
Then, the coupled equations for positive $\epsilon_0$ become
\begin{eqnarray}
\langle n_{AB}\rangle_{\pi/2}&\sim&\frac{\tilde{\Gamma}_{AB}}{2\pi\epsilon_0}(1-\langle n_d\rangle_{\pi/2})^2\ll 1,\\
\langle n_d\rangle_{\pi/2}&\sim&\frac{\Gamma_d}{2\pi\epsilon_0}(1-\langle n_{AB}\rangle_{\pi/2})^2\ll 1.
\end{eqnarray}
The solution of these coupled equations is
\begin{eqnarray}
\langle n_{AB}\rangle_{\pi/2}&\sim&\frac{\tilde{\Gamma}_{AB}}{2\pi\epsilon_0}\left(1-\frac{\Gamma_d}{\pi\epsilon_0} \right),\\
\langle n_d\rangle_{\pi/2}&\sim&\frac{\Gamma_d}{2\pi\epsilon_0}\left(1-\frac{\tilde{\Gamma}_{AB}}{\pi\epsilon_0} \right),
\end{eqnarray}
both of which are much smaller than $1$.

Next, we evaluate the AB modulation amplitude $\delta$ of $\langle n_{AB}\rangle$. For positive $\epsilon_0$, similar approach as in the previous discussion, we have
\begin{equation}
\langle n_{AB}\rangle\sim\frac{\Gamma_d}{2\pi\epsilon_0}(1-\langle n_{AB}\rangle_{\pi/2}-\delta\cos\phi)\ll 1.
\end{equation}
Since the AB phase modulation amplitude of $\langle n_d\rangle$ is much smaller than $1$, we neglect this dependence and hence
\begin{eqnarray}
\langle n_{AB}\rangle&\sim&\frac{1}{\pi}\int_{-\infty}^0\frac{d\epsilon}{\hbar}\frac{\frac{\tilde{\Gamma}_{AB}}{2}(1-\langle n_d\rangle_{\pi/2})^2}{\left(\frac{\epsilon-\epsilon_0}{\hbar}+\frac{\gamma_a}{2}\cos\phi \right)^2+\left[\frac{\tilde{\Gamma}_{AB}}{2}(1-\langle n_d\rangle_{\pi/2}) \right]^2}\nonumber\\
&\sim&\langle n_{AB}\rangle_{\pi/2}\left(1+\frac{\hbar\gamma_a}{2\epsilon_0}\cos\phi \right),
\end{eqnarray}
where we also assumed that $\epsilon_0\gg\gamma_a$. Therefore, we have
\begin{equation}
\delta\sim\langle n_{AB}\rangle_{\pi/2}\frac{\hbar\gamma_a}{2\epsilon_0}=\frac{\hbar^2\tilde{\Gamma}_{AB}\gamma_a}{\pi{\epsilon_0}^2}.
\end{equation}

Similar procedure as for $\epsilon_0>0$, we have
\begin{eqnarray}
\langle n_{AB}\rangle&\sim&\langle n_{AB}\rangle_{\pi/2}\left[1+\frac{\alpha\delta}{1-\langle n_d\rangle_{\pi/2}}\cos\phi-\frac{\frac{\hbar\gamma_a}{2\epsilon_0}+\frac{\alpha\delta}{1-\langle n_d\rangle_{\pi/2}}}{1-\frac{\tilde{\Gamma}_{AB}}{2\pi\epsilon_0}(1-\langle n_d\rangle_{\pi/2})}\cos\phi \right],
\end{eqnarray}
which should be equal to $\langle n_{AB}\rangle_{\pi/2}+\delta\cos\phi$. Here we introduced $\alpha\equiv\frac{\gamma_a}{2}(1-\langle n_d\rangle)$. Therefore, solving for $\delta$, we have
\begin{equation}
\delta\sim-\left(1+\frac{\lambda_a\sqrt{\lambda_a}}{\sqrt{\lambda_a}+\sqrt{\lambda_d}} \right)\frac{\hbar\gamma_a}{2\epsilon_0},
\end{equation}
where $\lambda_a\equiv\tilde{\Gamma}_{AB}/(2\pi|\epsilon_0|)$ and $\lambda_d\equiv\Gamma_d/(2\pi|\epsilon_0|)$. The required condition $|\delta|\ll1-\langle n_{AB}\rangle_{\pi/2}$, $\langle n_{AB}\rangle_{\pi/2}$ may be satisfied for $\tilde{\Gamma}_{AB}\sim\Gamma_d$.

Putting these results, the visibility becomes for positive $\epsilon_0$,
\begin{equation}
\eta\sim \frac{2\hbar^2\tilde{\Gamma}_{AB}\gamma_a}{\pi}\cdot\frac{1}{{\epsilon_0}^2}.
\end{equation}

Similarly, for negative $\epsilon_0$, we obtain
\begin{equation}
\eta\sim\frac{2(1+\lambda_a\langle n_{AB}\rangle_{\pi/2})}{1-\langle n_{AB}\rangle_{\pi/2}}\frac{\hbar\gamma_a}{2}\cdot\frac{1}{|\epsilon_0|}.
\end{equation}

These behaviors are consistent with the numerical results as shown in Fig. 4(c) and (d).

\section{QD energy dependence of visibility near Fermi level\label{ad-energy}}
Using the relation (\ref{population-relation}), at the limit of $V_C\to\infty$, we have
\begin{equation}
\langle n_{AB}\rangle=\frac{1}{\pi}\left\{\frac{\pi}{2}-(1-\langle n_d\rangle)\tan^{-1}\left[\frac{2(\epsilon_0-\alpha)}{(1-\langle n_d\rangle)\tilde{\Gamma}_{AB}} \right] \right\},
\end{equation}
where $\alpha$ is defined as $\alpha\equiv\frac{\gamma_a}{2}(1-\langle n_d\rangle)\cos\phi$ with $\gamma_a=\sqrt{\Gamma_S\Gamma_D\mathcal{T}_r}$. We assume that $\epsilon_0-\alpha\ll\tilde{\Gamma}_{AB}$ and use the approximation $\tan^{-1}x\simeq x$ for $x\ll 1$,
\begin{equation}
\langle n_{AB}\rangle\simeq\frac{1}{2}-\frac{2}{\pi\tilde{\Gamma}_{AB}}\left\{\epsilon_0-\frac{\gamma_a}{2}(1-\langle n_d\rangle)\cos\phi \right\}.
\end{equation}
Similarly, we obtain
\begin{equation}
\langle n_d\rangle\simeq\frac{1}{2}-\frac{2\epsilon_0}{\pi\Gamma_d}.
\end{equation}
Here we assumed that $\epsilon_0\ll\Gamma_d$. Thus, we obtain
\begin{equation}
\langle n_{AB}\rangle=\frac{1}{2}-\frac{2}{\pi\tilde{\Gamma}_{AB}}\left\{\epsilon_0-\frac{\gamma_a}{2}\left(\frac{1}{2}+\frac{2\epsilon_0}{\pi\Gamma_d} \right)\cos\phi \right\}.
\end{equation}
Using above result, from the definition of the visibility (\ref{visibility-def}), the visibility is given by
\begin{equation}
\eta\simeq\frac{32\hbar\gamma_a(\pi\hbar\tilde{\Gamma}_{AB})^3}{\left[(\hbar\gamma_a)^2-(\pi\hbar\tilde{\Gamma}_{AB})^2 \right]^2(\hbar\Gamma_d)^2}{\epsilon_0}^2.
\end{equation}
Therefore, the visibility shows the power-law behavior of ${\epsilon_0}^2$ near $\epsilon_0=0$.

\end{document}